\documentclass[competition,nonblindrev]{informs3-competition} 

\OneAndAHalfSpacedXI 
\usepackage{caption}
\usepackage{multirow}
\usepackage{booktabs}
\usepackage{array}
\usepackage{endnotes}
\let\footnote=\endnote

\TheoremsNumberedThrough     
\ECRepeatTheorems

\EquationsNumberedThrough    

\begin{document}
\RUNAUTHOR{Jiacheng Lu}

\RUNTITLE{Multi-modal\&Feedback Model for MVPP}

\TITLE{Multi-modal and Metadata Capture Model for Micro Video Popularity Prediction}

\ARTICLEAUTHORS{%
\AUTHOR{Jiacheng Lu, Mingyuan Xiao, Weijian Wang, Yuxin Du, Zhengze Wu, Cheng Hua}
\AFF{1954 Huashan Road, Shanghai, China, 200030, \EMAIL{cheng.hua@sjtu.edu.cn}
} 
} 

\ABSTRACT{
As short videos have become the primary form of content consumption across various industries, accurately predicting their popularity has become key to enhancing user engagement and optimizing business strategies. This report presents a solution for the 2024 INFORMS Data Mining Challenge, focusing on our developed 3M model (Multi-modal and Metadata Capture Model), which is a multi-modal popularity prediction model. The 3M model integrates video, audio, descriptions, and metadata to fully explore the multidimensional information of short videos. We employ a retriever-based method to retrieve relevant instances from a multi-modal memory bank, filtering similar videos based on visual, acoustic, and text-based features for prediction. Additionally, we apply a random masking method combined with a semi-supervised model for incomplete multi-modalities to leverage the metadata of videos. Ultimately, we use a network to synthesize both approaches, significantly improving the accuracy of predictions. Compared to traditional tag-based algorithms, our model outperforms existing methods on the validation set, showing a notable increase in prediction accuracy. Our research not only offers a new perspective on understanding the drivers of short video popularity but also provides valuable data support for identifying market opportunities, optimizing advertising strategies, and enhancing content creation. We believe that the innovative methodology proposed in this report provides practical tools and valuable insights for professionals in the field of short video popularity prediction, helping them effectively address future challenges.

}

\maketitle

\section{Introduction}

The widespread adoption of portable devices has significantly contributed to the success of micro video platforms like TikTok. These devices enable users to effortlessly share their experiences, opinions, and thoughts in various formats, such as text, images, audio, and video. The resulting increase in user engagement has given rise to an important research field: micro video popularity prediction (MVPP).\par

We recognize that current approaches to predicting the popularity of short videos often rely on tags rather than the videos themselves and fail to fully utilize their propagation context. This limitation means that existing methods are not effectively leveraging available data. To better apply multi-modal data and propagation context information, we have developed a new model applying an architecture of cross-modal attention combined with a semi-supervised model for incomplete modal estimation to predict video popularity.\par

3M is designed to make better use of the diverse data associated with short videos. It begins by using a retrieval system to find relevant micro videos from a multi-modal database, filtering content based on all available information, including visual, acoustic, and textual features, and using large models to extract features for comparison with existing samples. At the same time, we also create semi-supervised multi-modal model and obtain an additional correction of predicted values. By combining these two elements, our model has achieved good results in the task of short video popularity prediction.\par

\section{Methodology}

\begin{figure}
    \centering    \includegraphics[width=1\linewidth]{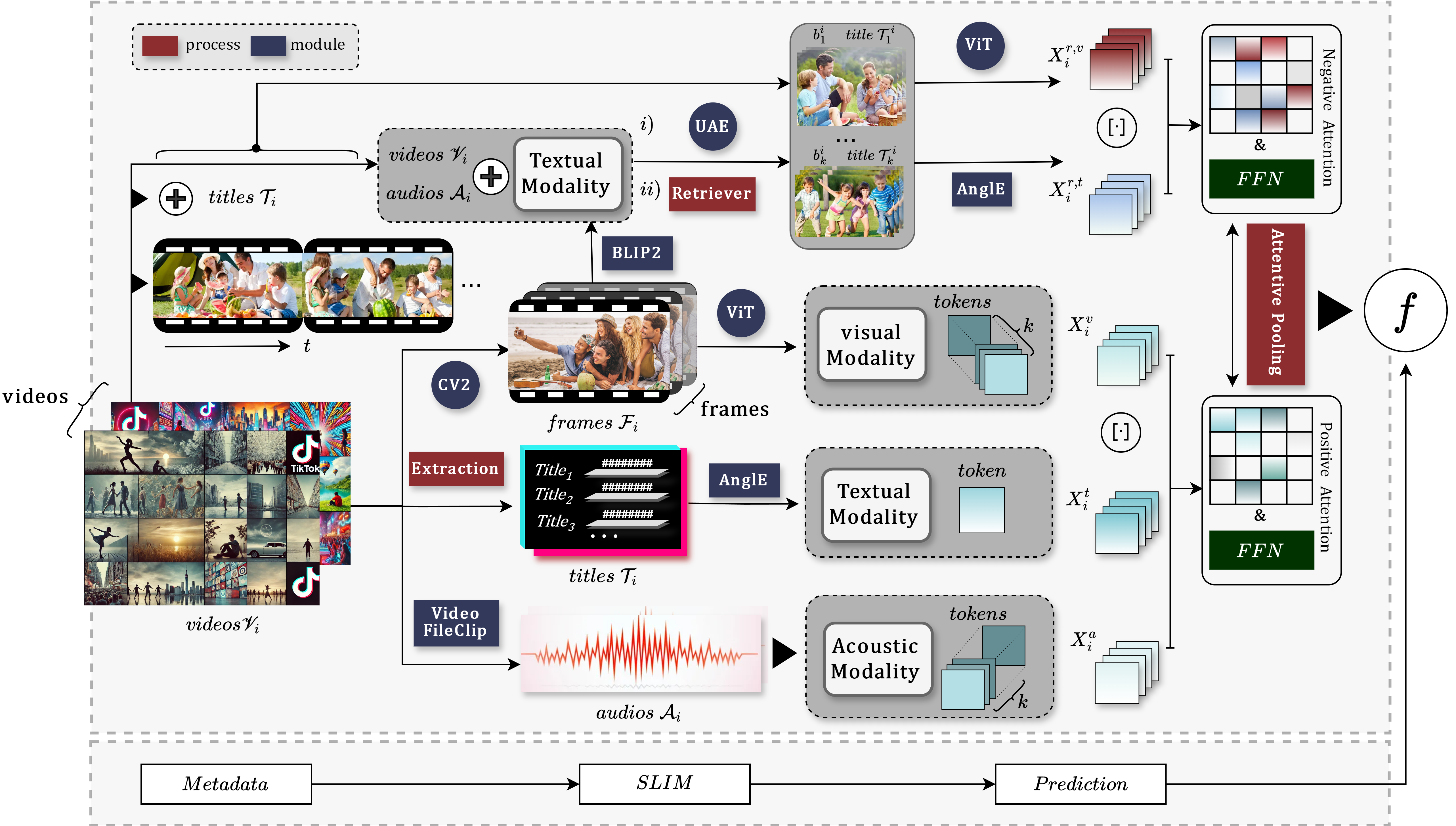}
    \begin{center}
        \caption{multi-modal information extraction module}
    \end{center}
    \label{fig1}
\end{figure}

Our methodological framework consists of two core modules: multi-modal information extraction module and metadata capture module. The video information extraction module utilizes multi-modal information processing and database detection. By leveraging a Large Language Model (LLM), we distill key insights from video content and generate descriptive annotations. Additionally, the LLM enables similarity-based search to retrieve related videos, while advanced architectures like the Vision Transformer (ViT) integrate diverse video modalities for comprehensive analysis. Subsequently, a cross-attention model explores the complex interactions between the retrieved video clusters and the target video.\par

\begin{figure}
    \centering    \includegraphics[width=1\linewidth]{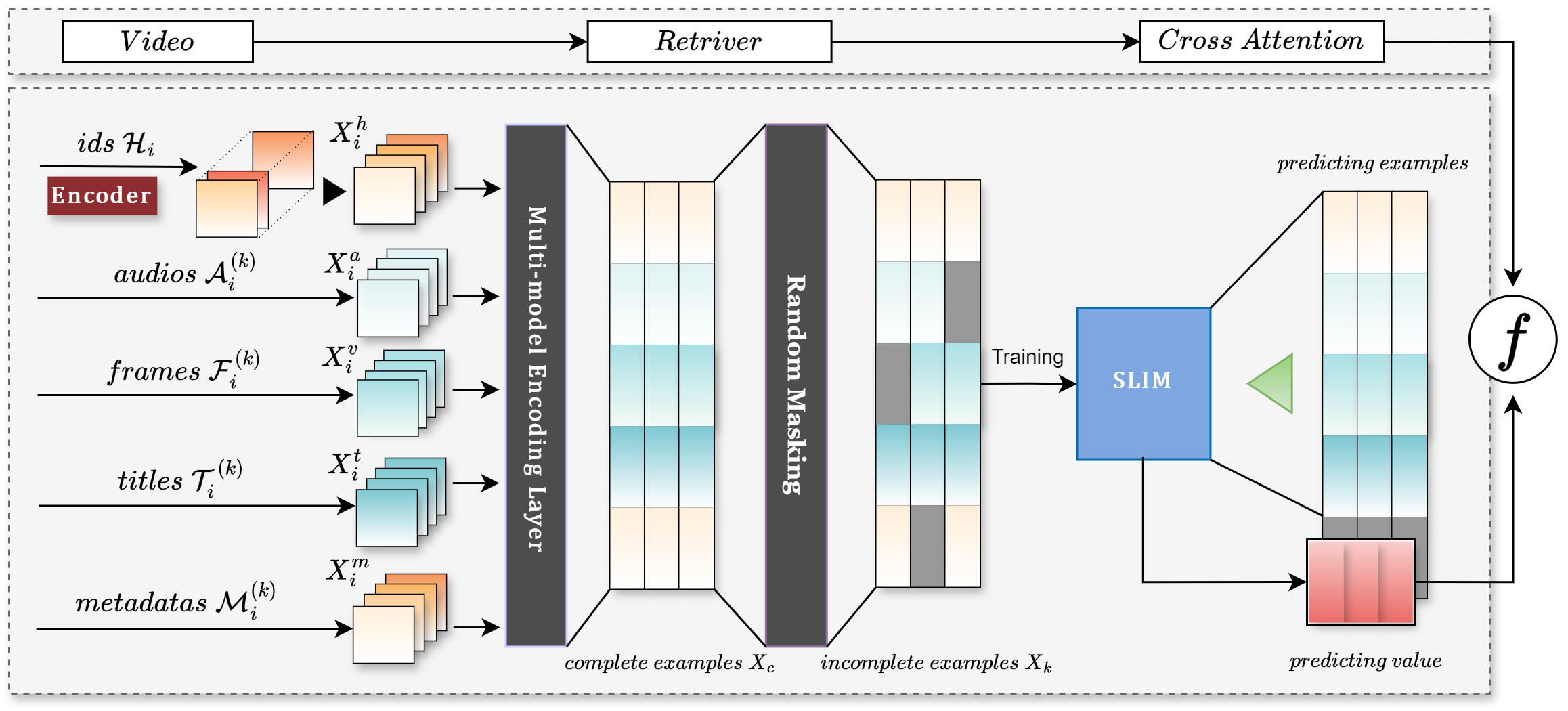}
    \begin{center}
        \caption{metadata capture module}
    \end{center}
    \label{fig2}
\end{figure}

In parallel, we attempt to introduce metadata to fully explore the multidimensional information of short videos. since metadata, which is the value our model predicts, is inherently missing in the target samples being predicted, we first employ a multi-modal dataset with random masking to obtain incomplete multi-modal group information. We then train the model using a semi-supervised multi-modal learning method tailored for missing multi-modalities, generating a robust multi-modal completion prediction model that can handle any partial feature missing. In this way, we can also attempt to generate a set of predicted values as supplemental correction in the final testing dataset, standing for the number of hearts, comments, shares and plays for each video.\par

In the final phase, an integrated network synthesizes the results from both modules, offering predictive insights into the video's potential to trend on social media platforms. This two-part model covers and utilizes as much relevant information from short videos as possible and generates a certain synergistic effect, with a prediction accuracy higher than that of the two sub-models.\par

\subsection{Results}

We check the 2203 training videos provided initially and found that only 1438 videos can be played correctly and completely. Then We discard the damaged videos, construct a new dataset with the remaining videos, and take 80\% of it for the training set and 20\% for the validation set. To evaluate the performance, we set two key metrics, $MSE$(Mean Square Error) and $PLCC$(Pearson Linear Correlation Coefficient).\par

After apply our methodology on the whole training set and get the evaluation metric, we attempt to train respectively on each author, since we found there are only 15 authors in the dataset. The model is respectively trained on training videos from a specific author and validated on validation videos from the same author. It may be considered as a trick, but this method indeed works and improves the performance of videos from each author. To further improve the model and avoid the issue of insufficient training data for certain authors, we combined the two aforementioned methods, comparing the performance between models trained on whole training set and on training set of a specific author, and choosing a better one. As a result, the enhanced model outperform the origin one noticeably, which is shown in Table\ref{Results}. We note the model trained comprehensively as \textbf{C}, the model trained respectively as \textbf{R}, the enhanced model combining better ones as \textbf{E}.\par

To evaluate the superiority of the model, we conduct experiments with 3 competitive baselines, using three methods \textbf{C}, \textbf{R} and \textbf{E} mentioned above. Our experimental baseline model includes: TMALL \cite{TMALL},  CBAN \cite{CBAN}, MASSL\cite{MASSL}. The performance of various baseline models and our author-enhanced prediction model on the validation dataset is presented in Table(\ref{Results}). \par

Finally, we apply our model on 258 testing videos, make predictions of share, heart, comment, play respectively for each author's videos. \par

\begin{table}[h]
\centering
\caption{MSE and PLCC values for different models and conditions}
\label{Results}
\begin{tabular}{>{\centering\arraybackslash}p{1.4cm}lcccccccc}
\toprule
& & \multicolumn{2}{c}{HEART} & \multicolumn{2}{c}{SHARE} & \multicolumn{2}{c}{COMMENT} & \multicolumn{2}{c}{PLAY} \\
& & MSE & PLCC & MSE & PLCC & MSE & PLCC & MSE & PLCC \\
\toprule
\multirow{3}{*}{OURS} & R & 6.4e+08 & \textbf{0.975} & 121113.7 & 0.983 & 80066.3 & 0.989 & 6.7e+10 & 0.961 \\
                      & C & 6.3e+08 & 0.975 & 148391.4 & 0.980 & 38979.9 & 0.986 & 6.8e+10 & 0.953 \\
                      & E & \textbf{5.3e+08} & 0.962 & \textbf{120510.4} & \textbf{0.981} & \textbf{30417.6} & \textbf{0.989} & \textbf{5.5e+10} & \textbf{0.967} \\
\midrule
\multirow{3}{*}{TMALL} & R & 8.4e+08 & 0.950 & 141113.8 & 0.969 & 100067.1 & 0.969 & 7.7e+10 & 0.940 \\
                      & C & 8.1e+08 & 0.950 & 168391.2 & 0.961 & 79890.5 & 0.956 & 7.8e+10 & 0.937 \\
                      & E & 7.3e+08 & 0.955 & 140510.6 & 0.963 & 50148.1 & 0.969 & 6.5e+10 & 0.949 \\
\midrule
\multirow{3}{*}{CBAN}  & R & 7.4e+08 & 0.963 & 131124.4 & 0.974 & 90166.2 & 0.979 & 7.2e+10 & 0.950 \\
                      & C & 7.4e+08 & 0.965 & 154391.7 & 0.969 & 58980.4 & 0.976 & 7.4e+10 & 0.944 \\
                      & E & 6.6e+08 & 0.965 & 132518.3 & 0.971 & 40414.7 & 0.978 & 6.0e+10 & 0.957 \\
\midrule
\multirow{3}{*}{MASSL} & R & 9.4e+08 & 0.926 & 151223.5 & 0.941 & 119263.3 & 0.948 & 8.7e+10 & 0.921 \\
                      & C & 9.3e+08 & 0.941 & 177380.8 & 0.940 & 89880.8 & 0.936 & 9.0e+10 & 0.913 \\
                      & E & 8.2e+08 & 0.945 & 150511.9 & 0.941 & 60897.9 & 0.950 & 7.5e+10 & 0.929 \\
\bottomrule
\end{tabular}
\end{table}

\section{Discussion}

We used the same but independent model for the four video metrics in the competition (shares, hearts, comments, plays), resulting in varied performances across these metrics. While our integrated model achieves satisfactory results overall, we see potential for improvement:\par

Exploring Metric Relationships: Time constraints prevented us from analyzing the internal relationships among the four metrics. Positive correlations likely exist, and their distributions may closely relate to video content—for example, landscape videos might receive more hearts, whereas opinion pieces garner more comments. By extracting structured features from video frames and audio spectra and combining the metrics, we could enhance the model's performance.\par

Incorporating Temporal Data: Including data such as comment timestamps or user behavior records could help build a more refined model of video dissemination. Sequential recommendation models like GRU4REC effectively capture how dissemination intensity changes over time. Although the competition data limited this integration, we believe incorporating these aspects could significantly improve the MVPP task.\par

\section{Conclusion}

In conclusion, our 3M Model has proven to be a robust approach for predicting micro-video popularity, offering significant improvements over conventional tag-based algorithms. By harnessing the power of multi-modal data and metadata, our model has achieved higher prediction accuracy, as evidenced by our validation set results.\par

While our model has demonstrated decent performance, there is still scope for refinement. Future work could focus on deeper analysis of the interplay between video metrics and the incorporation of sequential data to capture the dynamics of video popularity over time.\par

The 3M Model not only advances our understanding of micro-video popularity drivers but also provides actionable insights for strategic decision-making in content creation and advertising. We are confident that our research contributes meaningfully to the field and paves the way for further innovations in micro-video analytics.\par





\end{document}